\documentclass[a4paper,10pt]{article}
\usepackage{graphicx,eepic}
\usepackage{latexsym}
\usepackage{pb-diagram}
\usepackage{amsfonts, amsmath, amssymb, chapterbib, cite, dcolumn, epsfig,
rotating}

\def\b{\begin{eqnarray}}
\def\e{\end{eqnarray}}

\def\openone{\leavevmode\hbox{\small1\kern-0.355em\normalsize1}}

\newtheorem{theorem}{Theorem}

\allowdisplaybreaks

\begin{document}

\begin{center}

{\LARGE\textbf{The Camassa-Holm equation as a geodesic flow for the \\
$H^1$ right-invariant metric\\}} \vspace {10mm} \vspace{1mm}
\noindent

{\large \bf Adrian Constantin$^{a,\dag}$} and {\large \bf Rossen
I. Ivanov$^{a,\ast,}$}\footnote{On leave from the Institute for
Nuclear Research and Nuclear Energy, Bulgarian Academy of
Sciences, Sofia, Bulgaria.} \vskip 1cm \hskip -.3cm
\begin{tabular}{c}
\hskip-1cm $\phantom{R^R} ^{a}${\it School of Mathematics, Trinity
College Dublin,}
\\ {\it Dublin 2, Ireland} \\
\\{\it $^\dag$e-mail: adrian@maths.tcd.ie}
\\ {\it $^\ast$e-mail: ivanovr@maths.tcd.ie}
\\
\hskip-.8cm
\end{tabular}
\vskip1cm
\end{center}

\begin{abstract}
The fundamental role played by the Lie groups in mechanics, and
especially by the dual space of the Lie algebra of the group and the
coadjoint action are illustrated through the Camassa-Holm equation
(CH). In 1996 Misio\l ek observed that CH is a geodesic flow
equation on the group of diffeomorphisms, preserving the $H^1$
metric. This example is analogous to the Euler equations in
hydrodynamics, which describe geodesic flow for a right-invariant
metric on the infinite-dimensional group of diffeomorphisms
preserving the volume element of the domain of fluid flow and to the
Euler equations of rigid body whith a fixed point, describing
geodesics for a left-invariant metric on SO(3).

The momentum map and an explicit parametrization of the Virasoro
group, related to recently obtained solutions for the CH equation
are presented.
\end{abstract}



\section{Motion of a rigid body with a fixed point -- the SO(3) example}\label{aba:sec1}

Let us start with a very familiar example -- the Euler top. Consider
an orthogonal basis $\widetilde{e}_k(t)$, $k=1,2,3$, rotating about
a fixed basis $e_k$. Both bases share the same origin. We can think
about the moving frame as a rigid body, moving about the origin.

 The relation between the two frames is given by an
orthogonal transformation: $\widetilde{e}_k(t)=g_{kj}(t)e_j$, where
$g_{kj}=\widetilde{e}_k.e_j$,  $g^T=g^{-1}$, i.e. $g$ belongs to the
group $G \equiv SO(3)$, the corresponding algebra $\mathfrak{g}$
being
\begin{equation}
\mathfrak{g}\equiv so(3):  \qquad x\in \mathfrak{g} \Leftrightarrow
x=-x^T. \end{equation}

\begin{figure}
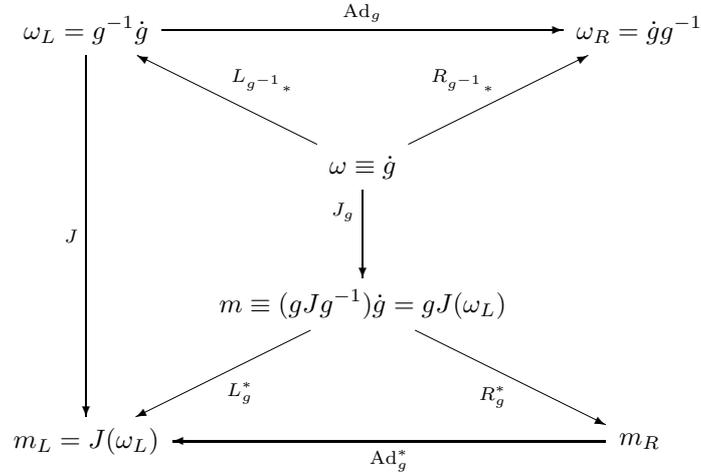

\[
\begin{diagram}
\node{\omega_L=g^{-1}\dot{g}}
      \arrow[2]{e,t}{\text{Ad}_g} \arrow[3]{s,l}{J}
   \node[2]{\omega_R=\dot{g}g^{-1}} 
\\
\node[2]{\omega\equiv\dot{g}}\arrow{nw,t}{{L_{g^{-1}}}_{*}}
\arrow{ne,t}{{R_{g^{-1}}}_{*}} \arrow[1]{s,l,1}{J_g}
\\
\node[2]{m\equiv(gJg^{-1})\dot{g}=gJ(\omega_L)} \arrow{se,b}{R_g^*}
\arrow{sw,b}{L_g^*}
\\
\node{m_L=J(\omega_L)} \node[2]{m_R}\arrow[2]{w,b}{\text{Ad}_g^*}
\end{diagram}
\]
\caption{Quantities and operators related to the so(3) algebra and
its dual (Euler top case).} \label{CH:fig1}
\end{figure}

\begin{figure}
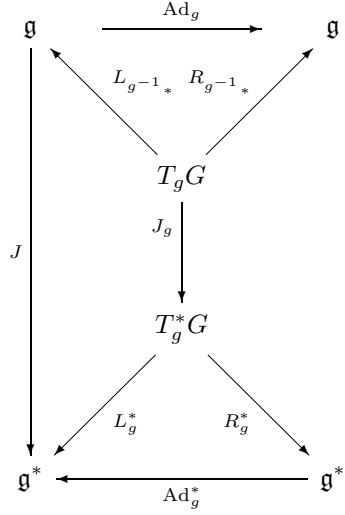

\[
\begin{diagram}
\node{\phantom{***}\mathfrak{g}\phantom{***}}
      \arrow[2]{e,t}{\text{Ad}_g} \arrow[3]{s,l}{J}
   \node[2]{\phantom{***}\mathfrak{g}\phantom{***}} 
\\
\node[2]{T_g G}\arrow{nw,t}{{L_{g^{-1}}}_{*}}
\arrow{ne,t}{{R_{g^{-1}}}_{*}} \arrow[1]{s,l,1}{J_g}
\\
\node[2]{\phantom{***}T_g^* G\phantom{***}} \arrow{se,b}{R_g^*}
\arrow{sw,b}{L_g^*}
\\
\node{\mathfrak{g}^*}
\node[2]{\mathfrak{g}^*}\arrow[2]{w,b}{\text{Ad}_g^*}
\end{diagram}
\]
\caption{Mappings between the spaces.} \label{CH:fig2}
\end{figure}

All quantities in the moving frame $\widetilde{e}_k(t)$ (related to
the body) will be marked by subscript '$L$', the ones in the fixed
basis - by subscript '$R$'. Let us take the following explicit
parametrization for the angular velocity:
\begin{equation}
\omega_L=\left(%
\begin{array}{ccc}
  0 & -\omega_3 & \omega_2 \\
  \omega_3 & 0 & -\omega_1 \\
  -\omega_2 & \omega_1 & 0 \\
\end{array}%
\right)\in \mathfrak{g}\quad \leftrightarrow \quad\vec{\omega}_L\equiv\left(%
\begin{array}{c}
  \omega_1 \\
  \omega_2 \\
  \omega_3 \\
\end{array}%
\right)\in \mathbb{R}^3 \label{eqomegaL} \end{equation}

The quantities related to the Euler top are schematically presented
at Fig. \ref{CH:fig1}, (the dot denotes the time
derivative)\cite{AK98,KM03,K04}. The identification between the
algebra $\mathfrak{g}$ and its dual is given by the {\it inertia
operator}, see Fig. \ref{CH:fig2}:
\begin{equation}m_L=J(\omega_L)\equiv A\omega_L+\omega_L A,\label{eqJ}
\end{equation} where $A=\text{diag}(a_1,a_2,a_3)$ is a constant symmetric
matrix.

The Hamiltonian is the kinetic energy
$H(m_L,\omega_L)=\frac{1}{2}\text{tr}(m_L\omega_L^T)$, given by a
left-invariant quadratic form: $H(m_L,\omega_L)=H(m,\omega)$.

This invariance by the virtue of Noether's Theorem leads to the
momentum conservation: $\frac{d}{dt}m_R=0$. This defines a momentum
map $TG\rightarrow\mathfrak{g}^*$, constant along the geodesics.
Furthermore, since $\omega_L=g^{-1}\dot{g}$ we have:
\begin{eqnarray}
m_L&=&\text{Ad}^*_g m_R=g^{-1}m_Rg, \qquad
\dot{m}_L=\text{ad}_{\omega_L}^*m_L=-[\omega_L,m_L].  \nonumber
\end{eqnarray} Finally we obtain the equations of motion (the Euler
top equations):
\begin{equation}
\frac{d}{dt}J(\omega_L)=[J(\omega_L),\omega_L] \qquad \text{or}
\qquad \dot{\omega}_1=\frac{a_2-a_3}{a_2+a_3}\omega_2\omega_3,
\qquad \text{etc.} \label{eqEuler} \end{equation}

\section{Camassa-Holm equation -- right invariant metric on the diffeomorphism group}
The construction described briefly in the previous section can be
easily generalized in cases where the Hamiltonian is a left- or
right-invariant bilinear form. Such an interesting example is the
Camassa-Holm (CH) equation \cite{FF81,CH93,J02}.  This geometric
interpretation of CH was noticed firstly by Misio\l ek \cite{M98}
and developed further by several other authors \cite{HMR98,
KM03,CK03,CKKT04,CK06,HM05}. Let us introduce the notation
$u(g(x))\equiv u\circ g$ and let us consider the $H^1$ Sobolev inner
product
\begin{equation}
H(u,v)\equiv \frac{1}{2}\int _{\mathcal{M}}(uv+u_xv_x)
d\mu(x),\qquad \text{ with} \qquad \mu(x)=x
\end{equation}
 The manifold $\mathcal{M}$ is
$\mathbb{S}^1$  or in the case when the class of smooth functions
vanishing rapidly at $\pm \infty$ is considered, we will allow
$\mathcal{M} \equiv \mathbb{R}$.


Suppose $g(x)\in G$, where $G\equiv \text{Diff}(\mathcal{M})$. Then
$H(u,v)=H(u\circ g, v\circ g)$ is a right-invariant $H^1$ metric.

Let us define $g(x,t)$ as
\begin{equation}
\dot{g}=u(g(x,t),t), \qquad g(x,0)=x, \qquad \text{i.e.} \qquad
\dot{g}=u\circ g \in T_g G;
\end{equation} $u=\dot{g}\circ g^{-1}={R_{g^{-1}}}_{*}\dot{g}\in
\mathfrak{g}$, where $\mathfrak{g}$ is $\text{Vect}(\mathcal{M})$.
Now we recall the following result:
\begin{theorem}[A. Kirillov, 1980] \cite{K81,K93}
The dual space of $\mathfrak{g}$ is a space of distributions but
the subspace of local functionals, called the regular dual
$\mathfrak{g}^*$,  is naturally identified with the space of
quadratic differentials $m(x)dx^2$ on $\mathcal{M}$. The pairing
is given for any vector field $u\partial_x\in
\text{Vect}(\mathcal{M})$ by

$\langle mdx^2, u\partial_x\rangle=\int_{\mathcal{M}}m(x)u(x)dx$

The coadjoint action coincides with the action of a diffeomorphism
on the quadratic differential:

$\text{Ad}_g^*:\quad mdx^2\mapsto m(g)g_x^2dx^2$
\end{theorem}
 If $m(x)>0$ for all $x\in \mathcal{M}$, then the square root
$\sqrt{m(x)dx^2}$ transforms under $G$ as a 1-form. This means that
$C=\int_{\mathcal{M}}\sqrt{m(x)}dx$ is a Casimir function, i.e. an
invariant of the coadjoint action.

\begin{figure}
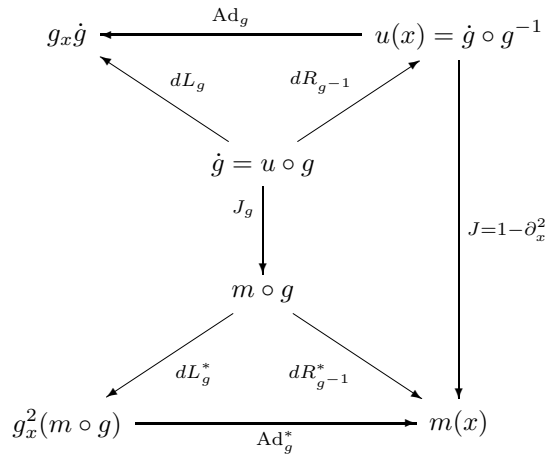

\[
\begin{diagram}
\node{g_x \dot{g}}
          \node[2]{u(x)=\dot{g}\circ g^{-1}} \arrow[2]{w,t}{\text{Ad}_g} \arrow[3]{s,r}{J=1-\partial_x^2}
          \\
\node[2]{\dot{g}=u\circ g} \arrow{ne,t}{dR_{g^{-1}}}
\arrow{nw,t}{dL_{g}} \arrow[1]{s,l,1}{J_g}
\\
\node[2]{m\circ g} \arrow{se,b}{dR_{g^{-1}}^*} \arrow{sw,b}{dL_g^*}
\\
\node{g_x^2(m\circ g)}\arrow[2]{e,b}{\text{Ad}_g^*} \node[2]{m(x)}
\end{diagram}
\]
\caption{Quantities related to the Camassa-Holm equation.}
\label{CH:fig3}
\end{figure}

 Let us now allow the above pairing to be the $H^1$
right-invariant metric, mentioned earlier. This is possible by
choosing the inertia operator $J=1-\partial_x^2$, i.e. by taking
$m=u-u_{xx}$, see Fig. \ref{CH:fig3}.  Again, for the Hamiltonian
$H=\frac{1}{2}\int _{\mathcal{M}}m u dx$, given by the $H^1$
right-invariant metric, Noether's Theorem yields \cite{CK03} the
conservation of $m_L\equiv g_x^2m(g(x,t),t)$, i.e.
$g_x^2m(g(x,t),t)=m(x,0)$.

We have a momentum map $TG\rightarrow\mathfrak{g}^*$, constant along
the geodesics:
\begin{equation}
0=\dot{m}_L=g_x^2(2u_xm+um_x+m_t)\circ g, \end{equation} iff $m$
satisfies the Camassa-Holm equation
\begin{equation}m_t+2u_xm+um_x=0.\end{equation}

Similarly to the Euler top (\ref{eqEuler}), CH can be written also
in a Hamiltonian form $\dot{m}=-\text{ad}_u^*m$. Indeed,
\begin{eqnarray}
\langle\text{ad}_{u\partial_x}^* mdx^2,v\partial_x\rangle&=&\langle
mdx^2,[u\partial_x,v\partial_x]\rangle=\int_{\mathcal{M}}m(u_xv-v_xu)dx=
\nonumber
\\ \int_{\mathcal{M}}v(2mu_x+um_x)dx&=&\langle(2mu_x+um_x)dx^2,v\partial_x\rangle, \nonumber \end{eqnarray}
i.e. $\text{ad}_{u}^*m=2u_xm+um_x$.


\section{Inverse Scattering for the CH equation}

In this section we consider the Camassa-Holm equation (CH) in the
form \begin{equation} u_{t}-u_{xxt}+2\omega
u_{x}+3uu_{x}-2u_{x}u_{xx}-uu_{xxx}=0, \label{CH}\end{equation}
which depends on an arbitrary parameter $\omega$ (which is not an
angular velocity!). The traveling wave solutions of (\ref{CH}) are
smooth solitons if $\omega > 0,$  and peaked solitons (peakons) if
$\omega = 0$ \cite{CH93,J02,CS1,CS2,L}.

If $\omega \ne 0$ the invariance group of the Hamiltonian is the
Virasoro group, $\text{Vir}=Diff(\mathbb{S}_1)\times \mathbb{R}$ and
the central extension of the corresponding Virasoro algebra is
proportional to $\omega$ \cite{CKKT04,I05}. Thus, for $\omega \ne
0,$ CH has various conformal properties \cite{I05}. CH is also
completely integrable, possesses bi-Hamiltonian form and infinite
sequence of conservation laws \cite{FF81,CH93,R02,CI,I06}. The Lax
pair is
\begin{eqnarray}
\Psi_{xx}&=&\Big(\frac{1}{4}+\lambda (m+\omega)\Big)\Psi
\label{Lax1}
\\
\Psi_{t}&=&\Big(\frac{1}{2\lambda}-u\Big)\Psi_{x}+\frac{u_{x}}{2}\Psi+\gamma\Psi\label{Lax2}
\end{eqnarray}
where $\gamma$ is an arbitrary constant (for a given eigenfunction).
CH is obtained from the compatibility condition
$\Psi_{xxt}=\Psi_{txx}$. Let us introduce a new spectral parameter
$k$ such that $\lambda(k)= -\frac{1}{\omega}\Big(
k^{2}+\frac{1}{4}\Big).$ From now on we consider the case where $m$
is a Schwartz class function, and $m(x,0)+\omega>0$. Then
$m(x,t)+\omega > 0$ for all $t\phantom{*}$ \cite{C01}.  The spectral
picture of (\ref{Lax1}) is\cite{C01}: continuous spectrum: $k$ --
real; discrete spectrum: finitely many points $k_{n}=\pm i\kappa
_{n}$, $n=1,\ldots,N$ where $\kappa_{n}$ is real and
$0<\kappa_{n}<1/2$. Eigenfunctions: for all real $k\neq 0$ a basis
in the space of solutions can be introduced, fixed by its asymptotic
when $x\rightarrow\infty$: $\psi(x,k)$ and $\bar{\psi}(x,k)$, such
that
\begin{equation}
\psi(x,k)=e^{-ikx}+o(1), \qquad x\rightarrow\infty.\end{equation}
Another basis can be introduced, fixed by its asymptotic when
$x\rightarrow -\infty$: $ \varphi(x,k)$ and $ \bar{\varphi}(x,k)$
such that
\begin{equation}
\varphi(x,k)=e^{-ikx}+o(1), \qquad x\rightarrow
-\infty.\end{equation} The relation between the two bases is
\begin{equation}\varphi(x,k)=a(k)\psi(x,k)+b(k)\bar{\psi}(x,k),\end{equation}
where\cite{CI}
\begin{equation}
|a(k)|^{2}-|b(k)|^{2}=1.\label{ab1}
\end{equation}
Further, one can define transmission and reflection coefficients:
$\mathcal{T}(k)=a^{-1}(k)$ and $\mathcal{R}(k)=b(k)/a(k)$
correspondingly. According to (\ref{ab1})
 \begin{equation}|\mathcal{T}(k)|^{2}+|\mathcal{R}(k)|^{2}=1.\nonumber \end{equation}

The entire information about these two coefficients is provided by
$\mathcal{R}(k)$ for $k>0$. It is sufficient to know
$\mathcal{R}(k)$ only on the half line $k>0$, since
$\bar{a}(k)=a(-k)$, $\bar{b}(k)=b(-k)$ and therefore
$\mathcal{R}(-k)=\bar{\mathcal{R}}(k)$. At the points of the
discrete spectrum, $a(k)$ has simple zeroes, $\varphi$ and
$\bar{\psi}$ are linearly dependent:
$\varphi(x,i\kappa_n)=b_n\bar{\psi}(x,-i\kappa_n).$ In other words,
the discrete spectrum is simple with eigenfunctions
$\varphi^{(n)}(x)\equiv \varphi(x,i\kappa_n)$. The asymptotic
behavior of $\varphi^{(n)}$ is
\begin{eqnarray}
\varphi^{(n)}(x)&=&e^{\kappa_n x}+o(e^{\kappa_n x}), \qquad
x\rightarrow -\infty;\nonumber\\\varphi^{(n)}(x)&=&b_n e^{-\kappa_n
x}+o(e^{-\kappa_n x}), \qquad x\rightarrow \infty. \label{phi_n}
\end{eqnarray} The sign of $b_n$ obviously depends on the number of
the zeroes of $\varphi^{(n)}$. Suppose that
$0<\kappa_{1}<\kappa_{2}<\ldots<\kappa_{N}<1/2$. Then from the
oscillation theorem for the Sturm-Liouville problem $\varphi^{(n)}$
has exactly $n-1$ zeroes, i.e. $ b_n= (-1)^{n-1}|b_n|$.

The set
\begin{equation}
\mathcal{S}\equiv\{ \mathcal{R}(k)\quad (k>0),\quad \kappa_n,\quad
|b_n|,\quad n=1,\ldots N\} \end{equation} is called {\it scattering
data}. The time evolution of the scattering data can be obtained
from (\ref{Lax2}) with the choice $\gamma=\frac{i k}{2\lambda}$ for
the eigenfunction $\varphi(k,x)$ and $x\rightarrow \infty$:
$\dot{a}(k,t)=0$, $ \dot{b}(k,t)= \frac{i k}{\lambda }b(k,t)$, or
\begin{equation} a(k,t)=a(k,0), \qquad b(k,t)=b(k,0)\exp\Big({\frac{i
k}{\lambda }t}\Big); \end{equation} In other words, $a(k)$ is
independent on $t$ and can serve as a generating function of the
conservation laws \cite{CI}.

 The time evolution of the data on the discrete spectrum is obtain as follows:
$i\kappa_n$ are zeroes of $a(k)$, which does not depend on $t$, and
therefore $\dot {\kappa}_n=0$. From (\ref{Lax2}) and (\ref{phi_n})
in a similar fashion \begin{equation} \dot{b}_n=\frac{4\omega
\kappa_n}{1-4\kappa_n^2}b_n. \qquad b_n(t)=b_n (0)\exp
\Big(\frac{4\omega \kappa_n}{1-4\kappa_n^2}t\Big ).\label{b_n}
\end{equation}

\section{Soliton solutions and the diffeomorphisms}

The inverse scattering is simplified in the important case of the
so-called reflectionless potentials, when the scattering data is
confined to the case $\mathcal{R}(k)=0$ for all real $k$. This class
of potentials corresponds to the $N$- soliton solutions of the CH
equation. In this case $b(k)=0$ and $|a(k)|=1$ and $ia'(i\kappa_p)$
is real:
\begin{equation}
ia'(i\kappa_p) = \frac{1}{2\kappa_p}e^{\alpha\kappa_p}\prod _{n\neq
p}\frac{\kappa_p-\kappa_n}{\kappa_p+\kappa_n},\quad\text{where}\quad
\alpha = \sum
_{n=1}^{N}\ln\Big(\frac{1+2\kappa_n}{1-2\kappa_n}\Big)^2.\nonumber
\end{equation}
Thus, $ia'(i\kappa_p)$ has the same sign as $b_n$, and therefore $
c_n\equiv \frac{b_n}{ia'(i\kappa_p)}>0. $ The time evolution of
$c_n$ is  $ c_n (t)=c_n(0)\exp\Big(\frac{4\omega
\kappa_n}{1-4\kappa_n^2}t\Big)$ in the view of (\ref{b_n}).

The $N$-soliton solution is \cite{CGI}
\begin{equation}
u(x,t)=\frac{1}{2}\int_0^{\infty}
\exp\Big(-|x-g(\xi,t)|\Big)p(\xi,t)d\xi-\omega,
\label{u}\end{equation} where $g(\xi,t)$, $p(\xi,t)$ can be
expressed through the scattering data as : \begin{eqnarray}
g(\xi,t)&\equiv&\ln
\int_0^{\xi}\Big(1-\sum_{n,p}\frac{c_n(t)\underline{\xi}^{-2\kappa_n}}{\kappa_n+1/2}A^{-1}_{np}[\underline{\xi},t]\Big)^{-2}d\underline{\xi},\label{g}\\
p(\xi,t)&=&\omega \xi^{-2}g_{\xi}^{-1}(\xi,t), \qquad \text{where} \\
A_{pn}[\xi,t]&\equiv&
\delta_{pn}+\frac{c_n(t)\xi^{-2\kappa_n}}{\kappa_p+\kappa_n}.\nonumber
\end{eqnarray}
Then the computation of $m=u-u_{xx}$ gives
\begin{equation}
m(x,t)=\int_{-\infty}
^{\infty}\delta(x-g(\xi,t))p(\xi,t)d\xi-\omega.
\label{m}\end{equation} From the CH equation $
m_t+um_x=-2(m+\omega)u_x$, (\ref{u}) and (\ref{m}) it follows
\begin{equation}
 \dot{g}(\xi,t)=\frac{1}{2}\int_0^{\infty}
e^{-|g(\xi,t)-g(\underline{\xi},t)|}p(\underline{\xi},t)d\underline{\xi}
 -\omega, \qquad
\dot{g}(\xi,t)=u(g(\xi,t),t), \nonumber\end{equation} therefore
$g(x,t)$ in (\ref{g}) is the diffeomorphism (Virasoro group element)
in the purely solitonic case. The situation when the condition
$m(x,0)+\omega>0$ on the initial data does not hold is more
complicated and requires separate analysis (if $m(x,0)+\omega$
changes sign there are infinitely many positive eigenvalues
accumulating at infinity and singularities might appear in finite
time \cite{CE98,C01}).


\section*{Acknowledgments}
A.C. acknowledges funding from the Science Foundation Ireland, Grant
04/BR6/M0042, R.I.I. acknowledges funding from the Irish Research
Council for Science, Engineering and Technology.

\bibliographystyle{ws-procs9x6}
\bibliography{ws-pro-sample}

\end{document}